\documentclass[oldversion]{aa}
\def\ltsima{$\;\buildrel < \over \sim \;$}
\def\simlt{\lower.5ex \hbox{\ltsima}}
\def\gtsima{$\;\buildrel > \over \sim \;$}
\def\simgt{\lower.5ex \hbox{\gtsima}}
\usepackage{graphicx}
\usepackage{txfonts}

\begin{document}

\title{Discovery of interstellar CF$^+$}
\author{David A. Neufeld\inst{1}, Peter Schilke\inst{2}, Karl M. Menten\inst{2},
Mark G. Wolfire\inst{3}, John H. Black\inst{4}, Fr\'ed\'eric Schuller\inst{2},
Holger~S.~P.~M\"uller\inst{2,5}, 
Sven Thorwirth\inst{2}, Rolf G\"usten\inst{2}, and Sabine Philipp\inst{2}}
\institute{Department of Physics and Astronomy, Johns Hopkins University,
3400 North Charles Street, Baltimore, MD 21218, USA
\and 
Max-Planck-Institut f\"ur Radioastronomie, Auf dem H\"ugel 69, 53121 Bonn, Germany 
\and 
Department of Astronomy, University of Maryland, College Park, MD 20742, USA 
\and
Chalmers University of Technology, Onsala Space Observatory, SE-43992 Onsala, Sweden 
\and
I.\ Physikalisches Institut, Universit\"at zu K\"oln, 50937 K\"oln, Germany }

\date{Received [date]; accepted [date]}

\abstract{We discuss the first astronomical detection of the 
CF$^+$ (fluoromethylidynium) ion, obtained by observations of
the $J=1-0$ (102.6~GHz), $J=2-1$ (205.2~GHz) and $J=3-2$ (307.7~GHz) rotational transitions toward the Orion Bar region. Our search for CF$^+$ -- carried out using the IRAM 30m and APEX 12m
telescopes -- was motivated by recent theoretical models that predict CF$^+$
abundances of $\rm few \times 10 ^{-10}$ in UV-irradiated molecular regions where C$^+$ is
present.  The CF$^+$ ion is produced by exothermic reactions of C$^+$ with HF.
Because fluorine atoms can react exothermically with H$_2$, HF is predicted to
be the dominant reservoir of fluorine, not only in well-shielded regions but
also in the surface layers of molecular clouds where the C$^+$ abundance is
large.  The observed CF$^+$ line intensities imply the presence of
CF$^+$ column densities $\ge 10^{12} \rm \, {cm}^{-2}$ over a region of size $\simgt 1^\prime$, in good
agreement with
theoretical predictions.  They provide support for our current theories of
interstellar fluorine chemistry, which suggest that hydrogen fluoride should
be ubiquitous in interstellar gas clouds and widely detectable in absorption by future satellite and airborne 
observatories. }

\authorrunning{Neufeld et al.}
\maketitle

\section{Introduction}

Of the $\sim 120$ interstellar molecules that have been detected to date\footnote{In addition, roughly two dozen molecules have been detected exclusively in circumstellar gas -- specifically around the source IRC+10216.  These molecules include species containing the elements Na, Mg, K, and Al, in addition to the elements represented
by molecules formed in the interstellar medium.} (e.g.\ M\"uller et al.\ 2005), more than 95$\%$ involve just six elements -- hydrogen, carbon,  nitrogen, oxygen, silicon, and sulfur -- all of which have moderately high abundances ($\ge 10^{-5}$ relative to hydrogen in the Sun).  Of the molecules that are formed in the interstellar medium, only four species containing other elements have been reported previously: HF (Neufeld et al. 1997), PN (Turner \& Bally 1987), HCl (Blake, Keene \& Phillips 1987) and FeO (Walmsley et al.\ 2002).  In the case of those elements of relatively low abundance (F, Cl, and P with solar abundances of $3 \times 10^{-8}$, $3 \times 10^{-7}$ and $3 \times 10^{-7}$ respectively with respect to H), the detected molecules are all stable species with large dissociation energies ($D^0_0 = 5.87$, 4.43 and 7.05 eV respectively for HF, HCl and PN\footnote{Derived from thermochemical data recommended in NIST Standard Reference Database Number 69 (a.k.a.\ the NIST Chemistry Web book), Eds. P.~J. Linstrom and W.~G. Mallard, March 2003, National Institute of Standards and Technology, Gaithersburg MD, 20899; available on-line at http://webbook.nist.gov}).  Not surprisingly, our understanding of the chemistries of interstellar fluorine-, chlorine-, phosphorus- and iron-bearing molecules is limited by the paucity of detected species that might help constrain theoretical models.

\begin{figure}
\includegraphics[scale=0.40,angle=-90]{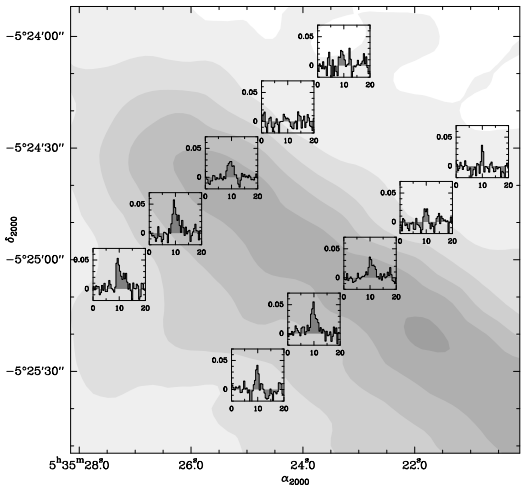}
\caption{CF$^+$ $J=1-0$ spectra observed at each of the positions in the Orion Bar, with the observed position for each observation indicated on a $\rm ^{13}CO$ $J=3-2$ map of the source (grayscale map, from Lis \& Schilke 2003).
For the strip centered on $(\alpha,\delta)$ = $\rm (05h\,35m\,22.8s$, $\rm -05^0\,25^\prime\,01^{\prime\prime})$ (J2000) (the ``Orion Bar (CO)'' position), the integrated CF$^+$ J = 1--0 antenna temperatures are $33 \pm 6$, $45 \pm 8$, $86 \pm 10$, $125 \pm 13$ and $57 \pm 8$ $\rm mK\,km\,s^{-1}$ from NW to SE  (quoted uncertainties are 
1 $\sigma$ statistical errors); for the strip centered on $\rm (05h\,35m\,25.3s$, 
$\rm -05^0\,24^\prime\,34^{\prime\prime})$ (J2000) (the ``Orion Bar (HCN)'' position), the corresponding values are $< 19.5$, $< 15.8$, $72 \pm 8$, $119 \pm 15$, and $140 \pm 30$ $\rm mK\,km\,s^{-1}$.}
\end{figure}

Notwithstanding this lack of observational constraints, the chemistry of fluorine-bearing molecules in the interstellar medium has been the subject of a recent theoretical study (Neufeld, Wolfire \& Schilke 2005a; hereafter NWS).
The principal conclusions of that investigation were:
1) that hydrogen fluoride is formed rapidly by the exothermic reaction of F atoms with H$_2$
and becomes the dominant reservoir of gas-phase F nuclei over a wide range of conditions; (2) that HF is abundant even close to UV-irradiated cloud surfaces where C$^+$ is the dominant reservoir of gas-phase carbon; and (3) that reaction of HF with C$^+$ can lead to potentially-measurable column densities of CF$^+$, a very stable molecule (dissociation energy, $D^0_0=7.71$~eV) that is isoelectronic with CO.

NWS showed that the chemistry of interstellar fluorine is qualitatively different from that of other elements because fluorine atoms can react exothermically with H$_2$, the dissociation energy of hydrogen fluoride being the largest of any neutral diatomic hydride and HF being the only such molecule more strongly bound than H$_2$.  Fluorine is therefore unique among the elements in having a neutral atom that reacts exothermically with H$_2$.

Motivated by the predictions of NWS, we used the 30m IRAM telescope and the 12m APEX\footnote{The Atacama Pathfinder Experiment (APEX)
is a collaboration of the Max-Planck-Institut f\"ur Radioastronomie, the European Southern Observatory, and the Onsala Space Observtory}
telescope to search for the $J=1-0$, $J=2-1$, and $J=3-2$ rotational transitions of CF$^+$ toward the Orion Bar, a well-studied photodissociation region (PDR) with an edge-on geometry favorable for the detection of molecular species of small abundance.  The observations are described in \S 2 below, and the results of those observations presented in \S 3.  A discussion follows in \S 4.

\section{Observations}

The $J=1-0$ rotational transition of CF$^+$ at 102.58748~GHz  (Plummer et al.\ 1986) was observed in two five-position strip maps in the Orion Bar, centered on the ``Orion Bar (CO)'' and ``Orion Bar (HCN)'' positions defined by Schilke et al.\ (2001; see our Figure 1 caption for coordinates).
The $J=2-1$ line at 205.17445~GHz and the $J=3-2$ line at 307.74438~GHz
were observed at the central ``Orion Bar (CO)'' and ``Orion Bar (HCN)'' positions for each strip.
 
The observations of CF$^+$ $J=1-0$ and $J=2-1$ were carried out
at the IRAM 30~m telescope, using the A100, B100, A230
and B230 receivers in position-switching mode, with an OFF-position
located 10$'$ away in azimuth. The backend employed was the VESPA correlation spectrometer; care
was taken to move the known backend defects away from the observed
lines.  The CF$^+$ $J=3-2$ transition was observed using the APEX 12m
telescope, located at an altitude of 5100~m on the Chajnantor plateau
in the Atacama desert of Chile, using the APEX-2a receiver.  The
observing mode was the same as at the 30~m, with the same
OFF-Position.  As backend, the MPIfR Fast Fourier Spectrometer was used. Calibration at
both telescopes is performed using a dual load scheme and an
atmospheric model. The beam sizes were 24$^{\prime\prime}$, 12$^{\prime\prime}$ and 21$^{\prime\prime}$ (HPBW),
respectively. Since the emission is extended, in the following antenna
temperature units are used.

\begin{figure}
\includegraphics[scale=0.34,angle=0]{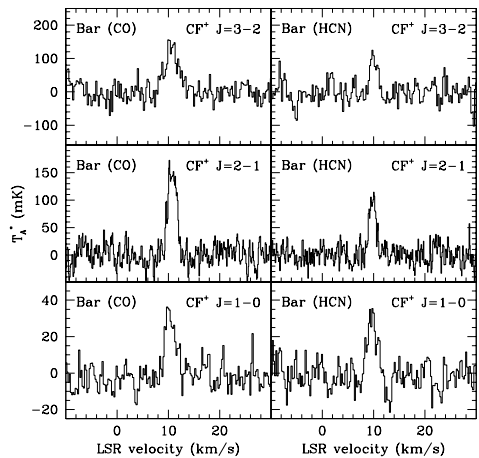}
\caption{CF$^+$ $J=1-0$, $J=2-1$ (IRAM 30~m) and $J=3-2$ (APEX 12m) spectra obtained toward the Orion Bar (CO) and Orion Bar (HCN) positions.
(Spectra of lower signal-to-noise ratio, obtained in shorter integrations toward the Orion Bar (CO) position, were published previously by Neufeld et al.\ 2005b.)}
\end{figure}

\begin{table}

\begin{minipage}[t]{\columnwidth}

\renewcommand{\footnoterule}{}  
\caption{Observations of the Orion Bar}
\centering
\begin{tabular}{lcc}
\hline \hline

&Orion Bar (CO) & Orion Bar (HCN) \\

\hline
&\multicolumn{2}{l}{Line strength\footnote{Errors given are 1 sigma statistical errors.  We estimate
the systematic uncertainties as $\pm 20\%$.} ($\int T_A^* dv$ in $\rm mK\,km\,s^{-1}$)} \\
$J=1-0$ & \phantom{0}$86 \pm 10$  &  \phantom{0}$72 \pm  8$\phantom{0} \\
$J=2-1$ & $337 \pm 13$ & $215 \pm 13$ \\
$J=3-2$ & $428 \pm 34$ & $183 \pm 26$ \\

&\multicolumn{2}{l}{Line centroid ($\rm km\,s^{-1}$ w.r.t. the LSR)}\\
$J=1-0$ & $10.5  \pm 0.15$ &  $\phantom{0}9.6  \pm 0.15$  \\
$J=2-1$ & $10.7  \pm 0.05$ &  $\phantom{0}9.8  \pm 0.05$  \\
$J=3-2$ & $10.6  \pm 0.1$\phantom{0}  &  $10.0 \pm 0.1$ \\

&\multicolumn{2}{l}{Line width ($\rm km\,s^{-1}$ FWHM)}\\
$J=1-0$ & $2.7\pm 0.4$ & $2.3 \pm 0.3$ \\
$J=2-1$ & $2.0\pm 0.1$ & $1.7 \pm 0.1$ \\
$J=3-2$ & $3.0\pm 0.3$ & $1.5 \pm 0.2$ \\

&\multicolumn{2}{l}{Column density (units of $\rm 10^{11}\,cm^{-2}$)} \\
$J=1$ & 3.7 & 3.1 \\
$J=2$ & 6.0 & 3.8 \\
$J=3$ & 4.7 & 2.0 \\
Estimated total & 19 & 11 \\
\hline \hline
\end{tabular}
\end{minipage}
\end{table}

\section{Results}

Figure 1 shows the CF$^+$ $J=1-0$ spectra observed at each of the positions in the Orion Bar, with the observed position for each observation indicated on a $^{13}$CO $J=3-2$ map of the source.  Emission in the $J=1-0$ line is evident at 8 of the 10 observed positions.  Like other optically-thin emission lines such as SiO $J=2-1$ (Schilke et al.\ 2001), the CF$^+$ $J=1-0$ line peaks southeast (i.e. further from the Trapezium stars) than $^{13}$CO $J=3-2$.  Rather than indicating that CF$^+$ and SiO are located further away from the ionization front than $^{13}$CO, this behavior reflects the different sensitivity to temperature and density - and the different optical depths - of the lines from these three species. In particular, the $^{13}$CO $J=3-2$ line is of moderate optical depth ($\tau > 1$), so its intensity probes the temperature at the $\tau \sim 1$ surface and is relatively insensitive to the total line-of-sight column density. 
Figure 2 shows the $J=1-0$, $J=2-1$ and $J=3-2$ spectra obtained at the central Orion Bar (CO) and Orion Bar (HCN) positions.  
Table 1 lists the integrated antenna temperature, line centroid, and line width for each observed transition at these two positions. 

\begin{figure}
\includegraphics[scale=0.26,angle=-90]{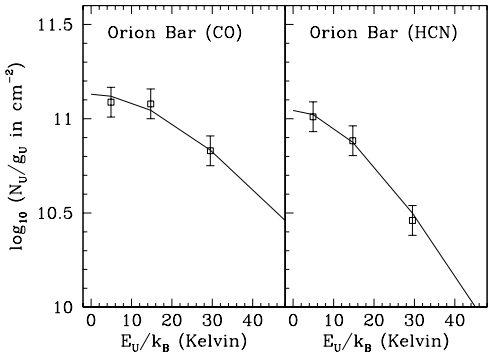}
\caption{Rotational diagram for CF$^+$.  The error bars include systematic uncertainties, which we estimate as $\pm 20\%$.
The solid curves show a fit to the data, in which the critical density for each transition is assumed proportional to the radiative decay rate.}
\end{figure}

Figure 3 shows a rotational diagram for the two positions toward which all three transitions have been observed,
computed for an assumed dipole moment of 1.07~D (Peterson et al.\ 1990).  Here, we assumed that the line emission is optically-thin, as suggested by the observed line strengths unless the covering factor of the emission is implausibly small.   The column densities for $J=1$, 2 and 3 are given in Table 1.  At both positions, any reasonable extrapolation of the rotational diagram implies that the $J=1$, $J=2$, and $J=3$ states are the three most highly-populated rotational states; thus the total column density we infer is only weakly dependent upon the extrapolation procedure adopted.    
The solid lines show the best fits to the rotational diagrams, which imply total CF$^+$ column densities of ${N(\rm CF^+}) = 1.9 \times 10^{12}\, \rm cm^{-2}$ and $1.1 \times 10^{12}\, \rm cm^{-2}$ respectively for the lines-of-sight to the Orion Bar (CO) and Orion Bar (HCN) positions.  Given our model predictions for the typical temperature in the CF$^+$ emitting region, and as suggested by the curvature of the rotational diagrams, we assumed here that the slope is mainly controlled by the density rather than the temperature, the former being insufficient for collisional deexcitation of $J=3$ to dominate spontaneous radiative decay.

\section{Discussion}

The results presented in Figure 1 and Table 1 indicate that CF$^+$ column densities $\sim 10^{12}\,\rm cm^{-2}$ are present over an extended region within the Orion Bar.  Such column densities are in good agreement with theoretical predictions (NWS), which predict CF$^+$ to be produced by the reaction sequence:
$$\rm H_2 + F \rightarrow HF + H\eqno(R1)$$
$$\rm HF + C^+ \rightarrow CF^+ + H \eqno(R2)$$
and destroyed primarily\footnote{Reaction with H$_2$ to form 
HCF$^+$, considered by 
Morino et al.\ (2000) to dominate the destruction of interstellar
CF$^+$, is in fact substantially endothermic$^2$ (by $\sim 2\, \rm eV$) and therefore of negligible importance (NWS).} by dissociative recombination
$$\rm CF^+ + e \rightarrow C + F \eqno(R3)$$

Because the reaction (R1) is exothermic and moderately rapid even at low temperature (Zhu et al.\ 2002), HF becomes the dominant reservoir of fluorine in the gas phase, even relatively close to cloud 
surfaces.  Beneath the surface of the UV-illuminated cloud, HF forms at 
precisely the point at which hydrogen becomes molecular, and long before carbon gets 
incorporated into CO (see NWS, Figure 2).  Thus there is a substantial area of overlap 
between $\rm C^+$ and HF where reaction (R2) can produce CF$^+$ abundances $\sim \rm few \times 10^{-10}$; here
CF$^+$ accounts for $\sim 1\%$ of the solar abundance of fluorine.

\begin{figure}
\includegraphics[scale=0.32,angle=0]{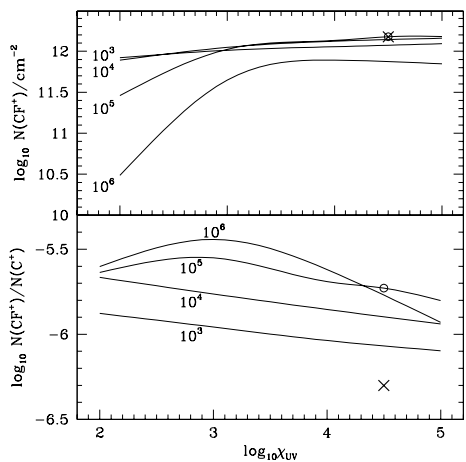}
\caption{a) Upper panel: CF$^+$ column density, $N({\rm CF^+})$, predicted by the NWS model, as a function of the normalized incident radiation field $\chi_{UV}$.
(b) Lower panel: corresponding $N({\rm CF^+})/N({\rm C^+})$ ratio.  Each curve is labelled with the assumed density, $n_H$. The crosses represent the mean observed values for the Orion Bar (CO) and Orion Bar (HCN) positions, along with the radiation field 
$\chi_{UV} = 3 \times 10^4$ estimated by Herrmann et al.\ (1997) for the Orion Bar, while the open circles represent the corresponding model prediction.}
\end{figure}

Figure 4a shows the CF$^+$ column density predicted by the NWS model, as a function of the gas density, $n_H$, and the incident radiation field $\chi_{UV}$ (again for the case of one-sided illumination at normal incidence).  The results apply to the case where fluorine depletion is estimated using the simple treatment given in NWS \S3.2  The predicted and observed results are in good agreement, although the remarkable agreement suggested by Figure 4a is undoubtedly fortuitous for several reasons:  (1) Figure 4a compares the predicted column densities for a photodissociation region (PDR) viewed face-on with the observed column densities in a nearly edge-on PDR (in which optically-thin lines are significantly limb-brightened); 
(2) the theoretical predictions are based upon several parameters that are substantially uncertain: e.g.\ the reaction rate coefficients for (R2) and (R3), and the gas-phase F abundance (see footnote 4).

The CF$^+$ column densities plotted in Figure 4a are found to track the corresponding C$^+$ column densities remarkably well, with the $N({\rm CF^+})/N({\rm C^+})$ column density ratio (Figure 4b) lying within a factor $\sim 2$ of $1.7 \times 10^{-6}$ for every case represented in Figure 4.
This behavior can be understood by 
considering the rates of CF$^+$ formation and destruction via reactions (R2) and (R3).  In equilibrium, we expect a ratio
$${n({\rm CF}^+) \over n({\rm C}^+)} = {k_2 \over k_3} {n({\rm HF}) \over n_e} = 4 \times 10^{-6}\biggl({ T \over 300\,{\rm K} } \biggr)^{0.35}
{n({\rm HF}) \over n_F} {n_C \over n_e}$$ 
where $k_2$ and $k_3$ are the rate coefficients given by NWS for (R2) and (R3), and $n_F/n_H = 1.8 \times 10^{-8}$ and $n_C/n_H = 1.6 \times 10^{-4}$ are the assumed gas-phase abundances of F and C.  In the region where the C$^+$ and CF$^+$ abundances are large, $n_e \sim n({\rm C^+}) \sim n_C$ and $n({\rm HF}) \sim n_F$. 

Kuiper Airborne Observatory (KAO) observations of the C$^+$ $^2P_{3/2} - ^2P_{1/2}$ (158$\mu$m) fine-structure line carried out toward the Orion Bar (Herrmann et al.\ 1997) permit us to compare our theoretical prediction for $N({\rm CF^+})/N({\rm C^+})$ with the measured value.  
Since both the C$^+$ and CF$^{+}$ transitions are optically-thin, such a comparison has the merit of eliminating any limb-brightening effects, all the observed lines being enhanced by the same factor.  Based upon their observations of the Orion Bar, Herrmann et al.\ inferred a line-of-sight C$^+$ column density of $3 \times 10^{18}\,\rm cm^{-2}$, averaged over a 55$^{\prime\prime}$ beam.  Given our estimates of $N({\rm CF}^+)$ at two separate positions along the midline of the bar, we may estimate the $N({\rm CF^+})/N({\rm C^+})$ ratio as $\sim 5 \times 10^{-7}$, a factor of $\sim 4$ smaller than the theoretical prediction.  This discrepancy may suggest (weakly, in light of all the uncertainties) that the F abundance assumed by NWS is too large and/or that the reaction (R2) proceeds less rapidly than was assumed.

Despite these uncertainties, our detection of CF$^+$ lends support to our theoretical model of fluorine chemistry.  In particular, the observed CF$^+$ column density argues strongly for a substantial region of overlap between C$^+$ and HF.  This, in turn, supports our contention that HF becomes abundant very close to cloud surfaces, and suggests that HF may prove a valuable tracer of molecular material in diffuse clouds.  Observations of HF and
CF$^+$ may also provide an important tool for studying the elemental 
abundance and nucleosynthetic origin of fluorine\footnote{The elemental abundance of fluorine is of particular interest because
the nucleosynthetic origin of fluorine remains controversial.  Fluorine is
produced in stars on the asymptotic giant branch (Jorissen, Smith, \&\
Lambert 1992; Werner, Rauch, \&\ Kruk 2005), but there is no consensus
on the relative contributions of other sources such as Wolf-Rayet
stars and type II supernovae (e.g. Renda et al. 2004; Palacios,,
Arnould, \&\ Meynet 2005). Because AGB stars exhibit a large
enhancement in the fluorine abundance (up to a factor $\sim 250$), it will be interesting to
search for CF$^+$ in circumstellar envelopes as well as in the
wind-blown bubbles surrounding Wolf-Rayet stars (Marston et al. 1999,
Marston 2001, Rizzo et al. 2003).}.
As noted by NWS, absorption line observations of the 1232 GHz HF $J=1-0$ transition will be possible with the Herschel Space Observatory and the 
Stratospheric Observatory for Infrared Astronomy; in addition, observations of HF at certain redshift ranges beyond $z=0.3$ are 
possible at ground-based observatories and will be greatly facilitated by ALMA.

\begin{acknowledgements}

We thank the staff of IRAM Granada, in particular Sergio Martin, for their help with the observations.
D.A.N.\ and M.G.W. gratefully acknowledge the support of grants NAG5-13114 and NNG05GD64G respectively from NASA's Long Term Space Astrophysics (LTSA) Research Program.  H.S.P.M. is supported by the Deutsche
Forschungsgemeinschaft (DFG) through grant SFB 494. 
IRAM is supported by INSU/CNRS (France), MPG (Germany), and IGN
(Spain). 

\end{acknowledgements}

\end{document}